\documentclass{article}

\usepackage{amssymb}

\def\toexp{\mathop{\rm exp}}
\newcommand{\Texp}{\toexp_{\leftarrow}}         

\newcommand{\bra}[1]{\langle #1 |}
\newcommand{\ket}[1]{| #1 \rangle}

\newcommand{\braOket}[3]{\langle #1 | #2 | #3 \rangle}

\newcommand{\bbraOket}[3]%
{\bigl\langle #1 \bigl| #2 \bigr| #3 \bigr\rangle}
\newcommand{\BbraOket}[3]%
{\Bigl\langle #1 \Bigl| #2 \Bigr| #3 \Bigr\rangle}
\newcommand{\Weak}{{\rm W}}
\newcommand{\Order}{\mathcal{O}}

\begin{document}

\title{Semiclassical theory of weak values}
\author{Atushi Tanaka\\
 Department of Physics, Tokyo Metropolitan University,\\
 Hachioji, Tokyo~192-0397, Japan }
\date{23 March 2002}
\maketitle
\begin{abstract}
 %
 Aharonov-Albert-Vaidman's weak values are investigated by a
 semiclassical method. 
 Examples of the semiclassical calculation that reproduces 
 ``anomalous'' weak values are shown.
 Furthermore, a complex extension of Ehrenfest's quantum-classical
 correspondence between quantum expectation values of the states with
 small quantum fluctuation, and classical dynamics, is shown.
 %
 %
 %
 %
\end{abstract}
%


\section{Introduction}

%
%
%
%

A standard approach of quantum theoretical description of the
interaction between a measurement apparatus and the system that is
subject to the measurement was introduced by von~Neumann.%
~\cite{Neumann:MGQ-1932}. 
Recently, it is revealed that the notion of 
{\em weak (expectation) values}%
~\cite{AAV:PRL-60-1351,AharonovVaidman:PRA-1990-41} emerges from
the von Neumann theory, in the limit that the influence of the
apparatus on the system is weak so as to avoid the collapse of
the state of the system.
The weak values and the conventional expectation values of quantum
theory coincide for the quantum ensembles that are specified by 
only the preselections (preparations) of initial states%
~\cite{AharonovVaidman:PRA-1990-41}.
However, once we consider the quantum ensembles that are specified not
only by the preselections of initial states, but also by the
postselections of finial states, the weak values depart from the
conventional expectation values%
~\cite{Aharonov:PLA-124-199,AAV:PRL-60-1351,AharonovVaidman:PRA-1990-41}.
In particular, the weak values can take ``anomalous'' values that lie
outside of the range of the eigenvalues of the corresponding
operators. 
More precisely, there are two kinds of anomalous weak values,
surprisingly large weak values of bounded operators, and
complex-valued weak values of Hermite operators.
On one hand, Aharonov, Albert and Vaidman showed an experimental setup,
in which weak measurements of a component of the spin of the
spin-$\frac{1}{2}$ particle can turn out to be 100~%
\cite{AAV:PRL-60-1351} (see also, \cite{Aharonov:PLA-124-199}).
Experimental studies confirms the surprisingly large weak values%
~\cite{RSH:PRL-66-1107,Parks:PRSLA-454-2297};
On the other hand, Aharonov, Albert and Vaidman~\cite{AAV:PRL-60-1351}
mentioned that the weak values can be a complex number.
This is also shown to be experimentally accessible%
~\cite{Parks:PRSLA-454-2297}, and moreover, 
to provide useful theoretical notions%
~\cite{Parks:JPA-33-2555,Wang:ComplexNelson,Steinberg:PRL-74-1995}.

This letter reports a study on weak values by a semiclassical method.
The usefulness of the semiclassical approach for the weak values was
anticipated by Berry's asymptotic
(i.e. semiclassical~\cite{Berry:RPP-35-315}) analysis of
superosillations~\cite{Berry:FTF-1994}. 
Although the anomalous weak values suggests us that the notion of weak 
values is completely foreign to classical concepts, the semiclassical
theory is shown to reproduce examples of anomalous weak values.
This is achieved by extending the classical trajectories into
complex-valued phase space.
%
It is well-known that the semiclassical method employs 
complex-valued classical trajectories to describe classically forbidden
phenomena,  e.~g.  tunneling~\cite{Tunneling} and nonadiabatic
transitions~\cite{ComplexNonadiabatic,Miller:JCP-56-1972}.
Furthermore, the recent investigations of the classically forbidden
processes in ``quantum chaos'' (quantum phenomena in the systems
whose classical counterpart exhibit chaos)~\cite{Gutzwiller:CCQM-1990}
revealed that the complex extension of classical trajectory is
indispensable~\cite{Adachi:AP-195-45,ChaoticTunnel}.

In the following, I remind the basic point of the weak measurements.
Consider the quantum ensemble that is specified not only by a
preselection (preparation) of an initial state but also a 
postselection of a final state: let the state vectors of the initial
and the final states be $\ket{\psi'}$ (at time $t'$) and $\bra{\psi''}$
(at time $t''(>t')$), respectively. The corresponding ``expectation
value'' of an observable $\hat{A}$ at an intermediate time $t$ ($t' \le
t \le t''$) is called a {\em weak value}  $\Weak (\hat{A}, t)$, whose
definition is  
\begin{equation}
 \label{eq:DefWeakValue}
 \Weak (\hat{A}, t) \equiv 
  \frac{\bbraOket{\psi''}{\hat{U}(t'',t)\hat{A}\hat{U}(t,t')}{\psi'}}%
  {\bbraOket{\psi''}{\hat{U}(t'',t')}{\psi'}}
\end{equation}
where $\hat{U}(t_1,t_0)$ is the time evolution operator during the time
interval~$[t_0, t_1]$~\cite{AAV:PRL-60-1351}.

The weak values are experimentally accessible by
a kind of von Neumann type apparatus~\cite{Neumann:MGQ-1932} whose
``pointer'' position has large quantum fluctuation. After the apparatus
weakly contact with the system and the system is succeedingly
postselected, we can gain the information of the corresponding weak
value by examining the statistics of the pointer. 
At each readout, the pointer value, which is obtained by the
standard quantum measurement that invokes wavepacket reduction, is very
noisy due to the initial fluctuation of the pointer and the weakness of
the interaction with the system. Accordingly we need to
accumulate the readouts of the pointer to make sense. 
The peak of the pointer position distribution determines the real part 
of the weak value. In addition, the imaginary part of the weak value is
determined by the peak of the momentum (i.e. the conjugate quantity of
the pointer position) distribution of the pointer. Thus both the real
and the imaginary parts of weak values are experimentally accessible%
~\cite{AAV:PRL-60-1351}.

The plan of this letter is as follows. In Sec.~\ref{sec:semilcassical}, I
develop a theory to evaluate weak values by a semiclassical method. 
Examples are shown in Sec.~\ref{sec:example}; We encounter 
two kinds of ``anomalous'' weak values, complex-valued weak values
(Sec.~\ref{subsec:complexEx}), and surprisingly large weak values
(Sec.~\ref{subsec:largeEx}).  
It is shown that they provide an ``anomalously-extended''
classical-quantum correspondence. 
I discuss the limitation of the present semiclassical argument
in Sec.~\ref{sec:limit}. Finally, Sec.~\ref{sec:SummaryAndOutlook}
summarizes this letter.

\section{Semiclassical evaluation of weak values}
\label{sec:semilcassical}

For brevity, I employ a one degree-of-freedom system, whose
position and momentum operators are $\hat{q}$ and $\hat{p}$, respectively.
Let us consider the quantum ensemble specified by an initial state
$\ket{\psi'}\equiv\ket{q'}$ at $t=t'$ and a final state
$\bra{\psi''}\equiv\bra{q''}$ at $t = t''$ ($>t'$), where $\ket{q}$ is 
the $\hat{q}$'s eigenvector whose eigenvalue is $q$.
In the evaluation of weak value $\Weak(\hat{A}, t)$, the following
generating functional is useful: 
\begin{equation}
 \label{eq:weakGF}
 Z(\zeta(\cdot), \hat{A}) \equiv 
  \langle\psi''|%
  \Texp\bigl\{-\frac{i}{\hbar}\int_{t'}^{t''} 
  \left(\hat{H} - \hat{A}\zeta(t)\right)dt
  \bigr\}%
  |\psi'\rangle
\end{equation}
where $\displaystyle\Texp(\cdot)$ is the time-ordered exponential and
$\hat{H}$ is the Hamiltonian of the system.
It is straightforward to show that
\begin{equation}
 \label{eq:weakValueByGF}
  \Weak (\hat{A}, t) =
  -\left. i\hbar\, 
    \frac{\delta\ln Z(\zeta(\cdot),\hat{A})}{\delta\zeta(t)}
  \right|_{\zeta(\cdot) = 0}
\end{equation}
holds. This is evaluated by a semiclassical method in the following.

Let $A(q,p)$ and $H(q,p)$ be the classical counterparts of the operator
$\hat{A}$ and the Hamiltonian $\hat{H}$ of the system, respectively. I
ignore the operator ordering problem, since it changes the result only
$\Order (\hbar)$, i.e., within the accuracy of the following
semiclassical argument.

In the evaluation of $Z(\zeta(\cdot),\hat{A})$~(\ref{eq:weakGF}),
I employ the semiclassical approximation that evaluates the Feynman path
integral representation of $Z$ by the stationary phase
method~\cite{Schulman:TAPI-1981}.
In order to carry out the semiclassical evaluations, I introduce an
important assumption: for infinitesimally small values of
$\zeta(\cdot)$, quantum interference between multiple classical
trajectories do not present in the semiclassical evaluation. Namely, the
semiclassical generating functional have a contribution only from one
classical trajectory $(q(t), p(t))$, which satisfies the Hamilton
equation with the classical Hamiltonian $H(q, p) - A(q, p)\zeta(t)$ and
the boundary condition $q(t') = q'$ and $q(t'') =q''$,
which are specified by the initial and the
final states $\ket{q'}$ and $\bra{q''}$, respectively%
\footnote{The Hamilton equation with the boundary condition $q(t') = q'$
and $q(t'') = q''$ can have multiple solutions. If $t''-t'$ is
small enough, it is proved that the boundary-value problem has
a unique solution. See, e.g., Ref.~\cite{Schulman:TAPI-1981}, Chap.~12.%
}. 
In other words, $Z$ is assumed to be in a single-term form%
\footnote{I omit Maslov's index~\cite{Gutzwiller:JMP-8-1979}, since this
is irrelevant to the present argument.} 
\begin{equation}
 \label{eq:ZqSingle}
 Z \simeq E \exp (i S /\hbar)
\end{equation}
where $E$ and $S$ are the amplitude factor and classical action,
respectively, 
\begin{eqnarray}
 \label{eq:Eq}
 E &\equiv& \frac{1}{\sqrt{2\pi \hbar\;\partial q''/\partial p'}}\\
 \label{eq:Sq}
 S &\equiv& \int_{t'}^{t''} 
  \{p(t) \dot{q}(t) - H(q(t), p(t)) \nonumber\\
 &{}& {}\quad\quad+ A(q(t), p(t)) \zeta(t)\}dt 
\end{eqnarray}
and $p' \equiv p(t')$~\cite{VanVleck:PNASUSA-14-178}.
The single-term condition~(\ref{eq:ZqSingle}) holds when $\hbar$ is
small or the time scale in question is short. The details are discussed
in Sec.~\ref{sec:limit}.
The single-term condition~(\ref{eq:ZqSingle}) implies
\begin{equation}
 \Weak (\hat{A}, t) 
  = \left. \frac{\delta S}{\delta\zeta(t)}\right|_{\zeta(\cdot) \equiv 0} 
 + \Order (\hbar).
\end{equation}
Applying~(\ref{eq:Sq}) to this, the main result is obtained:
\begin{equation}
 \label{eq:WKBWeakq}
  \Weak (\hat{A}, t)  = A(q(t), p(t)) + \Order (\hbar)
\end{equation}
where $\zeta(\cdot)\equiv 0$ is imposed on $(q(t), p(t))$.
The first term of eq.~(\ref{eq:WKBWeakq}) is a ``classical'' quantity:
It persists in the classical limit $\hbar\to0$ and is almost independent
of $\hbar$, in general.
The weak values are accordingly determined, with an error of
$\Order (\hbar)$, by the classical trajectory $(q(t), p(t))$ that
composes the semiclassical evaluation of the Feynman kernel 
$\braOket{\psi''}{\hat{U}(t'',t')}{\psi'} = Z(\zeta(\cdot)\equiv 0)$.

I emphasize the significance of the single-term
assumption~(\ref{eq:ZqSingle}), which optimize the shape of
semiclassical kernel, by using a simple wave~(\ref{eq:ZqSingle}) that is
composed by a single classical trajectory. The resultant
estimation~(\ref{eq:WKBWeakq}) accordingly exclude the effect of 
the quantum interference phenomena among multiple semiclassical
amplitudes. Namely, eq.~(\ref{eq:WKBWeakq}) establishes a correspondence
between a weak value and a single classical trajectory.

The generalization of the result~(\ref{eq:WKBWeakq}) to various
initial and final states (e.g. the eigenstates of the
momentum operator, and coherent states) can be obtained
straightforwardly with the help of the semiclassical
algebra~\cite{Miller:ACP-25-69,Weissman:JCP-76-4067}, as long as the 
single-term approximation~(cf. (\ref{eq:ZqSingle})) holds for
the corresponding semiclassical generating function $Z$.    

The ``variance'' of the weak value of $\hat{A}$ is 
\mbox{$\Weak (\{\hat{A} - \Weak (\hat{A})\}^2)$}~%
\cite{AharonovVaidman:PRA-1990-41}. According to the semiclassical
evaluation of weak values~(\ref{eq:WKBWeakq}), the ``weak variance'' is
$\Order (\hbar)$, when the single-term approximation~(\ref{eq:ZqSingle})
for the Feynman kernel holds.

\section{Examples: ``anomalous'' weak values}
\label{sec:example}

\subsection{Coherent state path integral:
An ``anomalous'' extension of quantum-classical correspondence into
complex-valued phase-space}
%
\label{subsec:complexEx}

Firstly, I show an example that the semiclassical theory above
reproduces complex-valued weak values. 
%
The semiclassical evaluation
(\ref{eq:WKBWeakq}) suggests that we encounter complex-valued weak
values for classically forbidden phenomena (e.g. tunneling phenomena
and nonadiabatic transitions). One of the simplest ways to investigate 
the classically forbidden phenomena is to study semiclassical coherent
state path integrals, which are generically composed of complex-valued
classical trajectory~\cite{SCPI}.
The present argument accordingly provides an extension of
quantum-classical correspondence into complex-valued phase-space.

I use the coherent states~\cite{KlauderSkagestram:CS-1985} that are
characterized with the help of a complex symplectic
transformation~\cite{KMS:GTIA-3-249,Weissman:JCP-76-4067}
\begin{equation}
 \label{eq:KMSTrans}
 \left[ \begin{array}{c} Q\\ P \end{array} \right]
 = 
  \left[
   \begin{array}{cc} 
    1/\sqrt{2}& -i/\sqrt{2}\\
    -i/\sqrt{2}& 1/\sqrt{2}\\
   \end{array} 
 \right]
 \left[ \begin{array}{c} q\\ p \end{array} \right]
\end{equation}
where $q$ and $p$ are the position and its canonical conjugate momentum
of the system, respectively. The quantized operators $\hat{Q}$ and 
$\hat{P}$ are creation and annihilation operators, respectively, of a
harmonic oscillator. 
The coherent state $\ket{q'p'}$ that is employed here is 
$\hat{P}$'s eigenstate whose eigenvalue is \mbox{$(p'-iq')/\sqrt2$}.

The semiclassical evaluation of $K(q''p''t'';q'p't') \equiv
\braOket{q''p''}{\hat{U}(t'',t')}{q'p'}$, which is the Feynman kernel in
the coherent state representation, is obtained by the stationary phase
evaluation of the coherent state path integral representation of
$K$~\cite{DaubechiesKlauder:JMP-26-2239}.
The boundary condition (at $t = t', t''$) of the classical trajectories
for semiclassical coherent state path integral is obtained by
Klauder~\cite{SCPI} 
\begin{eqnarray}
 \label{eq:CSKBC}
  \matrix{
 P(t') &=& P' \left(\equiv (p'-iq')/\sqrt{2}\right) \cr
 Q(t'') &=& Q'' \left(\equiv (q''-i p'')/\sqrt{2}\right)
    }
\end{eqnarray}
One way to explain the Klauder's boundary condition~(\ref{eq:CSKBC}) is
to remember the fact that $\ket{q'p'}$ is a right-eigenvector of $\hat{P}$
and $\bra{q''p''}$ is a left-eigenvector of $\hat{Q}$: The corresponding
eigenvalues determine the values of classical variables $P(t')$ and
$Q(t'')$. 
The classical trajectory $(q(t),p(t))$ during the time interval
$t'<t<t''$ is complex-valued in general, except the case that the
real-valued classical time evolution carries the point in the phase
space $(q',p')$ at time $t'$ to $(q'',p'')$ at time $t''$.

The single-term condition (cf. (\ref{eq:ZqSingle}))
holds for semiclassical Feynman kernel in the coherent state
representation, when the time interval $t'' - t'$ is small enough.
Accordingly the semiclassical estimation~(\ref{eq:WKBWeakq}) implies
\begin{equation}
 \label{eq:Wqp_sc}
  \left(\Weak(\hat{q}, t), \Weak(\hat{p}, t)\right) 
  = \left(q(t), p(t)\right) + \Order (\hbar).
\end{equation}
Namely, the weak values $\Weak(\hat{q}, t)$ and $\Weak(\hat{p}, t)$
approximately obey the classical equation of motion.
Furthermore, since $\left(q(t), p(t)\right)$ are generally
complex-valued as is explained above, so are the weak values
$\Weak(\hat{q}, t)$ and $\Weak(\hat{p}, t)$. 
%
%

The estimation~(\ref{eq:Wqp_sc}) provides an extension of a
correspondence between real-valued classical trajectories and the
expectation values of quantum systems.
The Ehrenfest theorem implies that the expectation values of quantum
system obeys the corresponding classical theory, as long as the quantum
fluctuation in a phase space representation is
small~\cite{EhrenfestTheorem}.  Although this 
concerns only for the real-valued, ``ordinary'' correspondence, the
semiclassical theory of weak values developed in this letter extends the
argument to the ``anomalous''-valued (i. e. complex-valued) trajectories.  
Note that the complex extension~(\ref{eq:Wqp_sc}) is carried out
without any discrimination between real- and complex-valued
trajectories.
This suggests that the distinction between real (``normal'') and complex
(``anomalous'') trajectories is only superficial in the framework of
the weak measurements.

Let us consider an example, the vanishing Hamiltonian $H=0$, which is
the simplest, yet nontrivial example in the studies of semiclassical coherent
state path integral~\cite{SCPI}. I remind that the semiclassical
evaluation for this system is exact. Let the initial  ($t =t'$) and
final ($t = t''$) states be $\ket{q'p'}$ and $\bra{q''p''}$,
respectively. During the time interval  $t' < t < t''$, 
the position and the momentum of the classical trajectory in
the semiclassical Feynman kernel are complex-valued in
general~\cite{SCPI}:  
\begin{eqnarray}
 q &=& \frac{1}{2}(q'' + q') - \frac{i}{2}(p''-p')\\
 p &=& \frac{1}{2}(p'' + p') + \frac{i}{2}(q''-q')
\end{eqnarray}
These are nothing but complex-valued weak values $\Weak(\hat{q}, t)$ and
$\Weak(\hat{p}, t)$ during the time interval  $t' < t < t''$.

\subsection{Spin coherent state path integral:
an example of anomalously large weak values}
\label{subsec:largeEx}

The weak values investigated in ref.~\cite{AAV:PRL-60-1351} 
for a spin-$\frac{1}{2}$ system can be reproduced
by~(\ref{eq:WKBWeakq}) the semiclassical theory of spin coherent state
path integral~\cite{SCPI}. 
Let $(\theta,\phi)$ be the orientation of the spin in the polar
coordinate, which parameterizes spin coherent states $\ket{\theta,\phi}$%
~\cite{KlauderSkagestram:CS-1985}. The classical
trajectory $(\theta(t),\phi(t))$ that composes the semiclassical
evaluation of a Feynman kernel
$\braOket{\theta'',\phi''}{\hat{U}(t'',t')}{\theta',\phi'}$ satisfies
Klauder's boundary 
condition~\cite{SCPI}  
\begin{equation}
 \label{eq:SpinKlauder}
  \matrix{
   e^{i \phi'} \tan (\theta'/2) &=&
    e^{i \phi (t')} \tan (\theta (t')/2) \cr
   e^{-i \phi''} \tan (\theta''/2) &=&
    e^{-i \phi (t'')} \tan (\theta (t'')/2)
    }
\end{equation}
as well as the classical equation of motion. Note that the classical
variables $\theta(t)$ and $\phi(t)$ are complex-valued in general.
The semiclassical weak values of the ensemble that is specified by an
initial state $\ket{\theta',\phi'}$ and a final state
$\bra{\theta'',\phi''}$ are obtained with the help of the semiclassical
formula~(\ref{eq:WKBWeakq}), if we do not encounter multiple classical
trajectories. 

For example, when the Hamiltonian of the system vanishes 
(i.e. $H\equiv 0$), the weak values of
the all components of the spin of the spin-$\frac{1}{2}$ particle
exactly agree with the values obtained by the semiclassical
evaluation~(\ref{eq:WKBWeakq}).
I demonstrate an example%
\footnote{For brevity, I simplified the example
studied in ref.~\cite{AAV:PRL-60-1351}.}%
$(\theta', \phi') = (2\alpha, 0)$, and
$(\theta'', \phi'') = (\frac{\pi}{2}, \pi)$, where $0<\alpha<\pi/2$.
The weak values, during the time interval $t'<t<t''$, are
\begin{eqnarray}
 \Weak(\hat{\sigma}_x) &=& -1\\
 \Weak(\hat{\sigma}_y) &=& 
  -i\; \frac{\cos\alpha + \sin\alpha}{\cos\alpha - \sin\alpha}\\
 \Weak(\hat{\sigma}_z) &=& 
  \frac{\cos\alpha + \sin\alpha}{\cos\alpha - \sin\alpha}
\end{eqnarray}
The value of $\Weak(\hat{\sigma}_x)$ is determined by the postselection;
$\Weak(\hat{\sigma}_y)$ is an example of complex-valued weak value;
$\Weak(\hat{\sigma}_z)$ can take surprisingly large values.

%
%
%

\section{Limitation of the semiclassical argument}
\label{sec:limit}

In order to obtain the semiclassical formula~(\ref{eq:WKBWeakq}), 
the single-term assumption~(cf.~(\ref{eq:ZqSingle})) is essential.
In this section, I discuss about the breakdown of this assumption.
%
%
I focus on the coherent state representation discussed in
Sec~\ref{subsec:complexEx}. 
Concerning to other pairs of initial and final states, the similar
phenomena occur (more precisely, see ref.~\cite{Littlejohn:JSP-68-7}).
When $t''-t'$ is small, the semiclassical Feynman kernel have a
significant contribution only from the single classical trajectory that
are placed around the real-valued classical trajectory. Since (phase
space) caustics \cite{Adachi:AP-195-45,Caustics},
which are bifurcation points of classical trajectories, exist far from
the real-valued trajectory, the influence from the caustics is small.
The single-term assumption accordingly holds. 
As $t''-t'$ become larger, several caustics approaches to the
real-valued trajectory~\cite{AT:Preprint-1999}. Consequently the
influence from the caustics to the Feynman kernel become significant to
produce the quantum interference phenomena between multiple classical
trajectories. Hence the single-term assumption breaks
down~\cite{Adachi:AP-195-45}.
Furthermore, the caustics induce the divergence of the semiclassical
amplitude factor (cf. eq.~(\ref{eq:Eq})). Such divergence induces the
divergence of $\Order (\hbar)$ contribution to
$\Weak(\hat{A})$~(\ref{eq:WKBWeakq}) as well as the divergence of the
semiclassical evaluation of weak variance 
$\Weak(\{\hat{A}-\Weak(\hat{A})\}^2)$.  
Although the semiclassical method itself does not breakdown even in the
presence of the quantum interference, the semiclassical expressions of
the weak values become complicated in general. 

In summary, at the breakdown of the single-term assumption, the
semiclassical evaluation encounters large fluctuations due to the
caustics. After the large fluctuations, the interference between the
multiple classical trajectories emerge.

\section{Summary}
\label{sec:SummaryAndOutlook}

The present argument establishes an intimate relationship between weak
values and classical trajectories that appear in the semiclassical
evaluations of Feynman kernels, when the quantum interference between
multiple classical trajectories are negligible. 
In particular, it is shown that the semiclassical theory has an ability
to reproduce complex-valued or surprisingly large, ``anomalous'', weak
values, with the help of the complex-valued classical trajectories.

\section*{Acknowledgments}
A part of this work is carried out at Institute of Physics,
University of Tsukuba, where this work is supported by University of
Tsukuba Research Projects.



\end{document}